\documentclass[modern]{aastex631}
\usepackage{amsmath,amssymb,graphicx}

\begin{document}

\title{Indistinguishability criterion and estimating the presence of biases}

\author{Alexandre Toubiana}
\affiliation{Max Planck Institute for Gravitationsphysik (Albert Einstein Institute) \\
Am M\"{u}hlenberg 1 \\ 14476 Potsdam, Germany}

\author{Jonathan R. Gair}
\affiliation{Max Planck Institute for Gravitationsphysik (Albert Einstein Institute) \\
Am M\"{u}hlenberg 1 \\ 14476 Potsdam, Germany}


%





\begin{abstract}
    In these notes, we comment on the standard indistinguishability criterion often used in the gravitational wave (GW) community to set accuracy requirements on waveforms~\citep{Flanagan:1997kp, Lindblom:2008cm, McWilliams:2010eq, Chatziioannou:2017tdw, Purrer:2019jcp}. Revisiting the hypotheses under which it is derived, we propose a correction to it. Moreover, we outline how the approach described in~\cite{Toubiana:2023cwr} in the context of tests of general relativity can be used for this same purpose. 
\end{abstract}

\section{Definitions}

In the following, we use $\theta$ to denote the parameters of a GW signal model, $n_p$ the total number of parameters and $h(\theta)$ the family of templates we use to analyse the GW data. We assume that we have a set of $n_d$ independent observations, which could be the strain in different detectors or the time-delay-interferometry (TDI) variables $A$, $E$ and $T$ \citep{Tinto:2004wu} in the case of the Laser Interferometer Space Antenna (LISA) \citep{Audley:2017drz}. In general, the strain $d$ is the sum of a GW signal $s_0$ and noise. Here, we work in the zero-noise approximation, taking $d=s_0$. The ``true'' GW signal $s_0$ need not belong to the family of model templates, $h(\theta)$, but we assume it depends on the same set of parameters, $\theta$. Denoting $\theta_0$ the parameters such that $s_0=s(\theta_0)$, we wish to derive a simple criterion to determine if when analysing the observed data with our family of templates $h$, we would obtain a biased estimate of $\theta_0$ due to systematic effects. 

We define the noise-weighted inner product between two data streams, $d_1$ and $d_2$, as: 
\begin{equation}
    (d_1|d_2)=4 {\mathcal Re} \left [ \int_0^{+\infty}\frac{d_1(f)d^*_2(f)}{S_n(f)}df \right ], \label{eq:inner_product}
\end{equation}
where $S_n(f)$ is the power spectral density (PSD). For a given choice of PSD, the SNR of a signal $s$ is defined as ${\rm SNR}=\sqrt{(s|s)}$. 
The posterior distribution of the source parameters, $\theta$, given datasets $d_1,...,d_{n_d}$ is given by Bayes' theorem:
\begin{equation}
    p(\theta|d_1,...,d_{n_d})=\frac{p(d_1,...,d_{n_d}|\theta)p(\theta)}{p(d_1,...,d_{n_d})},
\end{equation}
where $p(d_1,...,d_{n_d}|\theta)$ is the likelihood, which we denote by $\mathcal{L}(\theta)$ in the following, $p(\theta)$ is the prior and $p(d_1,...,d_{n_d})$ is the evidence. Throughout these notes, we assume the prior on $\theta$ to be flat and that its support contains the support of the likelihood. Thus, up to normalisation constants, which do not affect its shape, the posterior on $\theta$ is equal to the likelihood. 
Assuming the noise to be stationary, Gaussian and independent for each observed dataset, the likelihood reads:
\begin{equation}
   \mathcal{L}= p(d_1,...,d_{n_d}|\theta|)=\prod_{i}^{n_d}\exp \left [ -\frac{1}{2}(s_{0,i}-h_i(\theta)|s_{0,i}-h_i(\theta)) \right ].\label{eq:logl}
\end{equation}
%

The log-likelihood can then be written:
\begin{align}
    \ln\mathcal{L}&= \sum_{i}^{n_d} (s_{0,i}|h_i(\theta)) -\frac{1}{2} (s_{0,i}|s_{0,i})-\frac{1}{2} (h_i(\theta)|h_i(\theta)),  \\
    &= \sum_{i}^{n_d} (s_{0,i}|s_{0,i}) \sqrt{\frac{(h_i(\theta)|h_i(\theta))}{(s_{0,i}|s_{0,i})}} \left ( \frac{(s_{0,i}|h_i(\theta))}{\sqrt{(s_{0,i}|s_{0,i}) (h_i(\theta)|h_i(\theta))}} \right. \nonumber \\ &\hspace{6cm} \left.
     -\frac{1}{2} \sqrt{\frac{(s_{0,i}|s_{0,i})}{(h_i(\theta)|h_i(\theta))}} -\frac{1}{2} \sqrt{\frac{(h_i(\theta)|h_i(\theta))}{(s_{0,i}|s_{0,i})}} \right ).\label{eq:logl_exp}
\end{align}
We define the overlap between two data streams as: 
\begin{equation}
    \mathcal{O}(d_1,d_2)=\frac{\left( d_1 | d_2\right)}{\sqrt{\left( d_1 | d_1\right)\left( d_2 | d_2\right)}}.\label{eq:overlap}
\end{equation}
Then, assuming that the true signal and our recovered templates have similar loudness, i.e., $(s_{0,i}|s_{0,i})\simeq (h_i(\theta)|h_i(\theta))$\footnote{We assume this to hold in the region of the parameter space we would identify when performing parameter estimation.}, from Eq.~\ref{eq:logl_exp} we can write the log-likelihood as:
\begin{align}
    \ln\mathcal{L}(\theta)&= -\sum_i^{n_d} {\rm SNR_i}^2 (1-\mathcal{O}_i(s_0,h(\theta))), \\
    &= -{\rm SNR}^2_T (1-\mathcal{O}_T(s_0,h(\theta))). \label{eq:log_like_olap}
\end{align}
In the above equation, we have introduced the total SNR, ${\rm SNR}_T$, and the total overlap, $\mathcal{O}_T$, defined as:
 \begin{align}   
    {\rm SNR}^2_T&=\sum_i^{n_d} {\rm SNR_i}^2, \\
    \mathcal{O}_T(s_0,h(\theta))&=\sum_i^{n_d} \frac{{\rm SNR_i}^2}{{\rm SNR}_T^2} \mathcal{O}_i(s_0,h(\theta)).  
\end{align}
Next, we use the linear signal approximation \citep{Finn:1992wt} to write the likelihood in a Gaussian form. We denote by $\hat{\theta}$ the maximum likelihood point, and decompose the GW signal into a component within the template manifold and a component orthogonal to it (in the sense of the inner product defined in Eq.~\ref{eq:inner_product}) as: \mbox{$s_{0,i}=h_i(\hat{\theta})+\delta h_i$}, such that $\sum_{i=1}^{n_d} (\delta h_i|\partial_j h_i(\hat\theta))=0 \  \forall \  j$, where $\partial_j h = \partial h/\partial \theta^j$ as usual.
This yields:
\begin{equation}
    \mathcal{L}=\hat{\mathcal{L}} \exp \left[ -\frac{1}{2} (\theta-\hat{\theta})^t \left (\sum_i^{n_d}F^i\right ) (\theta-\hat{\theta}) \right ], 
     \label{eq:gauss_like}
\end{equation}
where the maximum likelihood, $\hat{\mathcal{L}}$, and the Fisher matrices, $F^i$, are given by:
\begin{align}
    \hat{\mathcal{L}}&=\exp \left [-\frac{1}{2} \sum_i^{n_d}(\delta h_i|\delta h_i) \right ], \label{eq:max_logl} \\
    F^i_{j_k}&=(\partial_j h_i|\partial_k h_i)_{|_{\hat{\theta}}}.  \label{eq:Fisher}
\end{align}

The Gaussian approximation is expected to hold for high enough SNRs. Note that its validity is independent of the overall quality of the fit, which is measured by $\hat{\mathcal{L}}$. 


\section{Indistinguishability criterion}

\subsection{Standard criterion}

The standard indistinguishability criterion is obtained by:
\begin{enumerate}
    \item assuming $\hat{\mathcal{L}}=1$,
    \item requiring that the log-likelihood at the true point $\theta_0$ is larger than $\exp[-n_p/2]$, the log-likelihood $1-\sigma$ away in all directions from $\hat{\theta}$. 
\end{enumerate}
From  Eqs.~\ref{eq:log_like_olap} and \ref{eq:gauss_like} these conditions translate into:
\begin{equation}
    1-\mathcal{O}_T(s_0,h(\theta_0))  \leq \frac{n_p}{2{\rm SNR}_T^2}. \label{eq:st_crit}
\end{equation}
Although very handy, this criterion tends to be too conservative and biases are not necessarily found when it is violated \citep{Purrer:2019jcp}. It can be readily improved upon by modifying the two hypotheses under which it was derived. 


\subsection{Revisiting the criterion}

The first one requires that the component of the GW signal orthogonal to the template family $h$ is null, see Eq.~\ref{eq:max_logl}, i.e., there exists some set of parameters for which our template can perfectly reproduce the signal. However, this is usually not the case. For instance, when recovering NR injections with \texttt{SEOBNRv5HM} templates including all harmonics in~\cite{Toubiana:2023cwr}, our maximum log-likelihoods were always significantly below 0 (see Fig.~19 of that paper), and since they scale with the ${\rm SNR}^2$, this becomes even more relevant as the SNR increases. This problem can be corrected by adding the actual maximum log-likelihood. We will discuss ways to estimate this in section~\ref{sec:prat_cons}.
%
%

\begin{figure}[!h]
\centering
 \includegraphics[width=\textwidth]{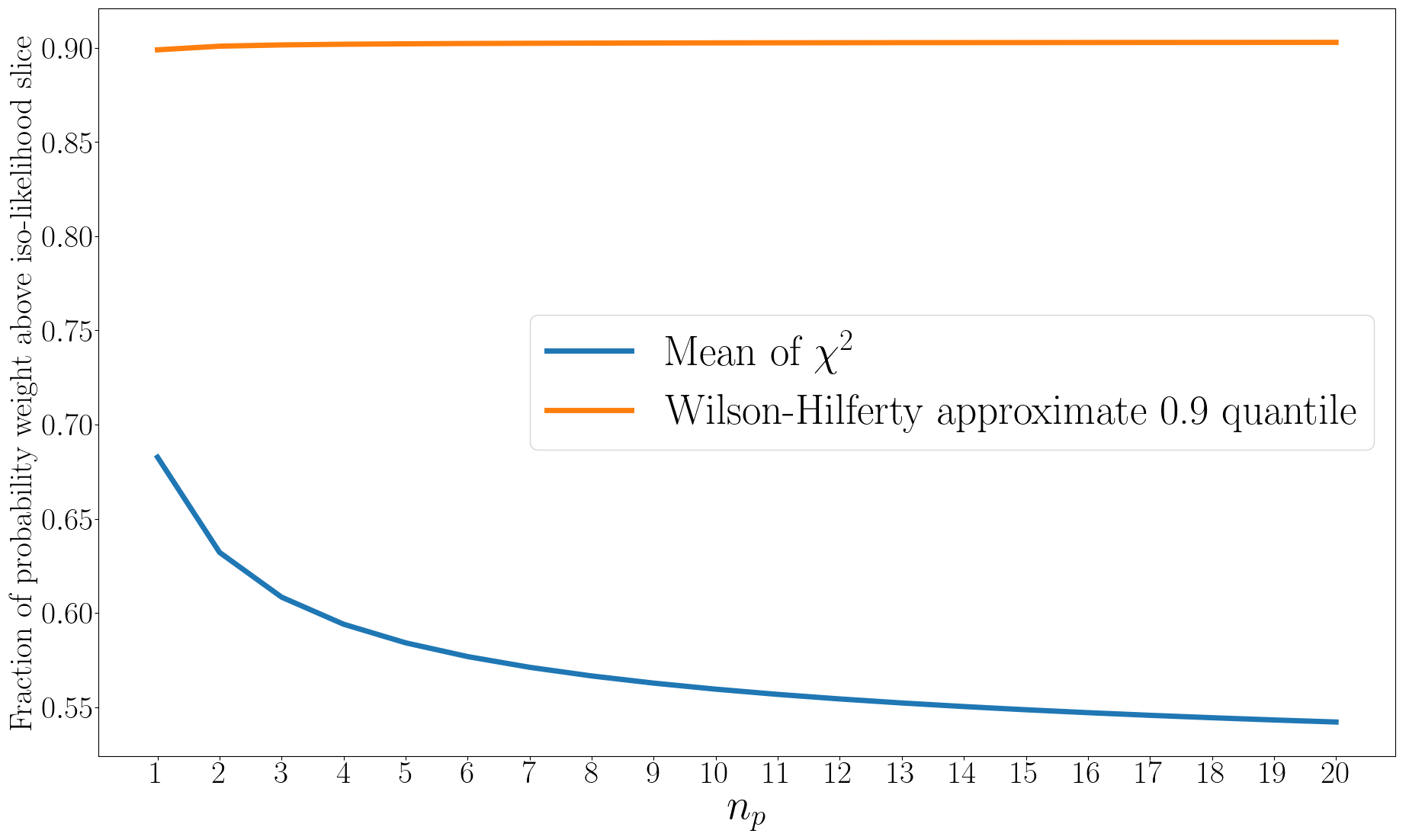}\\
 \centering
 \caption{Fraction of the probability weight of an $n_p$-dimensional Gaussian distribution for the part of the parameter space above a given iso-likelihood slice as a function of $n_p$. On a one hand, taking the mean value of the log-likelihood, this fraction varies substantially with $n_p$. On the other hand, the expression defined in Eq.~\ref{eq:app_q90_logl} is very stable and gives a good approximation to the $90\%$ quantile.}\label{fig:chi2_weights}
\end{figure}
The second hypothesis is motivated by the $1-\sigma$ rule for Gaussian distributions. Using the fact that $2(\ln\hat{\mathcal{L}}-\ln \mathcal{L})$ is distributed as a $\chi^2(n_p)$ distribution\footnote{This can readily be seen by transforming Eq.~\ref{eq:gauss_like} into the basis of eigenvectors of the total covariance matrix, which then clearly shows that $2(\ln\hat{\mathcal{L}}-\ln \mathcal{L})$ is the sum of $n_p$ normalised Gaussian random variables, and so follows a $\chi^2(n_p)$ distribution.}, we can reinterpret the second hypothesis as requiring that the log-likelihood at the true point is larger than the mean log-likelihood. However, as shown in Fig.~\ref{fig:chi2_weights}, for an $n_p-{\rm dimensional}$ Gaussian distribution, the fraction of the probability weight contained in the region with log-probability larger than the mean log-probability depends significantly on $n_p$. 
 It starts from $0.68$ for $n_p=1$, as expected for the $1-\sigma$ contour of a one-dimensional Gaussian, and tends slowly to 0.5\footnote{The 0.5 limit comes from the fact that as $n_p \rightarrow \infty$, the $\chi^2$ distribution approaches a Gaussian distribution with mean $n_p$ and variance $2n_p$. However, this convergence is very slow.}. 
 Better motivated lower limits for the log-likelihood at the true point, are the quantiles of the log-likelihood distribution, as also proposed in \cite{Baird:2012cu}. 
 Schematically, we want to take an iso-likelihood slice of the $n_p-{\rm dimensional}$ parameter space such that a specified fraction, $f$, 
 of the probability weight is contained above that slice. If the true point lies above that slice, it is contained in the $n_p-{\rm dimensional}$ $100 f \,\%$ confidence region. Denoting by $\mathcal{Q}_{f}(n_p)$ the $100 f\,\%$ quantile of the $\chi^2(n_p)$ distribution, the revisited criterion then reads:
\begin{equation}
    \ln \mathcal{L}(\theta_0) \geq  \ln \hat{\mathcal{L}}-\frac{\mathcal{Q}_{f}(n_p)}{2},\label{eq:crit_general}
\end{equation}

 To obtain a handy expression, we fix $f=0.9$ and use the Wilson–Hilferty transformation~\citep{b71582f9-54b0-3efd-95a7-a60d1005dd3b}\footnote{If $X$ follows a $\chi^2(n_p)$ distribution, then $\sqrt[3]{\frac{X}{n_p}}$ converges to a normal distribution with mean $1-\frac{2}{9n_p}$ and variance $\frac{2}{9n_p}$. Remarkably, this convergence is much faster than the one of $\chi^2$ to a normal distribution.}, and approximate $\ln \mathcal{L}_{0.9}$ as:
\begin{equation}
    \ln \mathcal{L}_{0.9}=\ln \hat{\mathcal{L}}-\frac{n_p \left ( 1-\frac{2}{9 n_p}+1.3\sqrt{\frac{2}{9n_p}} \right )^3 }{2}. \label{eq:app_q90_logl}
\end{equation}
The 1.3 factor is ad-hoc such that this approximates the $90\%$ quantile of the log-likelihood distribution, as can be seen in Fig.~\ref{fig:chi2_weights}. 
%
%
Finally, defining the total fitting factor ${\rm FF}_T$ as the maximised overlap over $\theta$, so that:
\begin{equation}
    \ln \hat{\mathcal{L}}= -{\rm SNR}^2_T (1-{\rm FF}_T(s_0,h)),
\end{equation}
we can write the revised indistinguishability criterion as: 
\begin{equation}
     1-\mathcal{O}_T(s_0,h(\theta_0)) \leq \frac{n_p \left ( 1-\frac{2}{9 n_p}+1.3\sqrt{\frac{2}{9n_p}} \right )^3 }{2{\rm SNR}_T^2}+1-{\rm FF}_T(s_0,h). \label{eq:crit}
\end{equation}

 \begin{figure}[!h]
\centering
 \includegraphics[width=\textwidth]{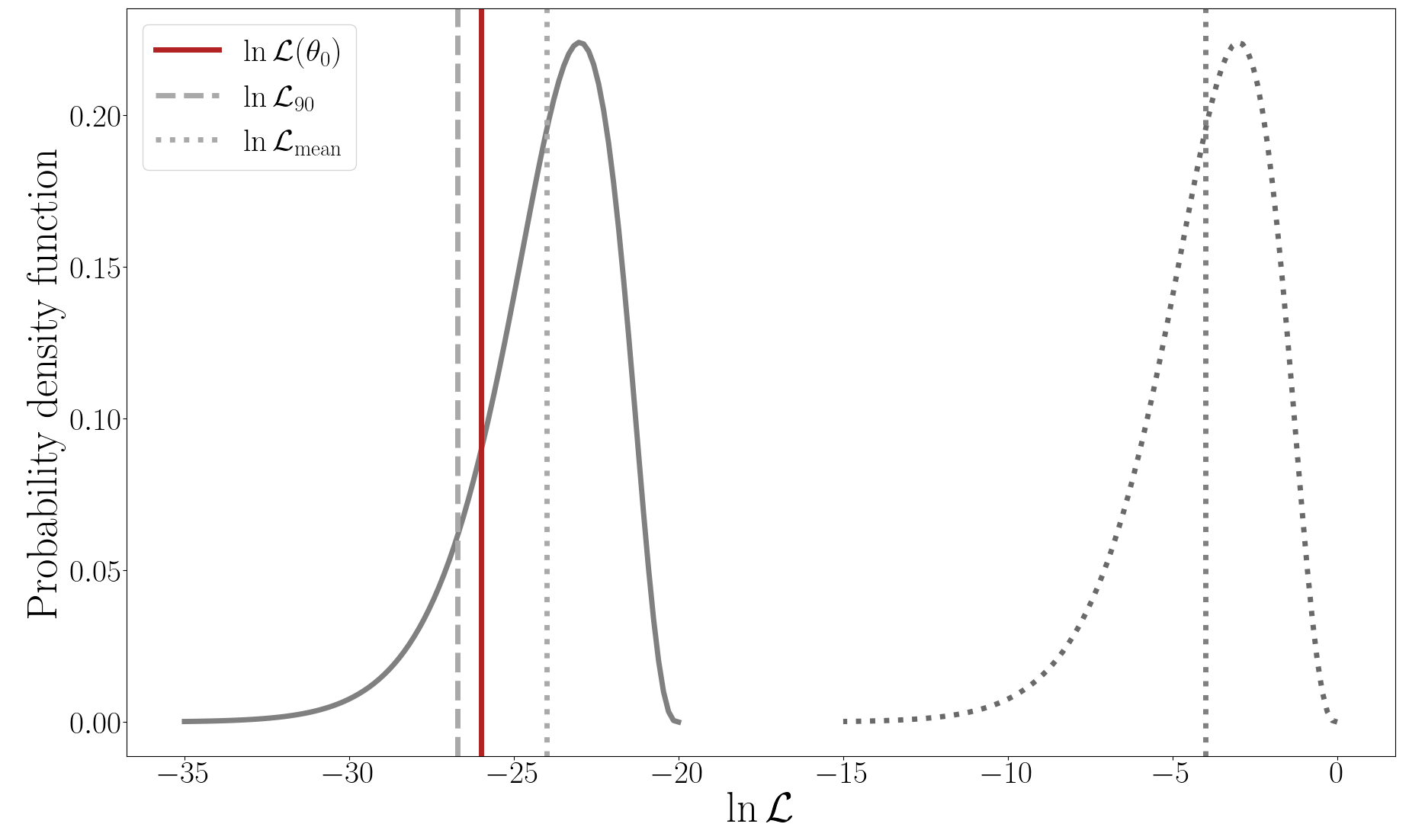}\\
 \centering
 \caption{Probability density function of log-likelihood values in a mock case with $n_p=8$ parameters, $\ln \mathcal{L}(\theta_0)=-26$ and $\ln \hat{\mathcal{L}}=-20$. The dotted curve assumes $\ln \hat{\mathcal{L}}=0$, as in the standard version of the indistinguishability criterion. The vertical dotted lines show the mean log-likelihood value, for both the true and the wrong value of $\hat{\mathcal{L}}$. In the standard formulation of the indistinguishability criterion, it is the lower limit for the log-likelihood at the true point to decide on the presence of biases. The dashed line shows the $90\%$ quantile of the true log-likelihood distribution. This plot illustrates why using the standard indistinguishability criterion, one would erroneously conclude that biases are expected, whereas our revisited criterion (Eq.~\ref{eq:crit}) corrects for this.}\label{fig:plot_criterion}
\end{figure}

In Fig.~\ref{fig:plot_criterion}, we illustrate why the standard indistinguishability criterion could erroneously indicate that we expect a bias, when our revisited criterion would not. The standard criterion makes reference to the idealised likelihood distribution of the true model (where $\ln \hat{\mathcal{L}}=0$), while the new criterion makes reference to the measured likelihood distribution. Moreover, the former requires that the log-likelihood at the true point is larger than the mean log-likelihood, which, in addition to not being statistically sound, tends to be too stringent. This illustration shows a hypothetical scenario in which the value of the log-likelihood at the true point lies well within the bulk of the distribution, above the $90\%$ quantile, indicating that the true point should lie in the $90\%$ confidence region, while the whole support of the log-likelihood distribution lies well below the idealised likelihood distribution reference by the standard criterion. Accounting for the corrections we propose is expected to decrease the discrepancies between the prediction of the indistinguishability criterion and full parameter estimation found in other studies, including ~\cite{Purrer:2019jcp}.\footnote{In that work the authors let the factor in the numerator of the right-hand side of Eq.~\ref{eq:st_crit} be free instead of setting it to $n_p$, and estimate it by setting an equality between the left and right-hand sides of Eq.~\ref{eq:st_crit} at the SNR where they find, from full from parameter estimation, systematic biases to equal statistical errors. In one case, they find this factor to be as large as $10^4$ for an SNR of 250, which could be obtained for a fitting factor of 0.92. This is plausible considering that they analyse numerical relativity waveforms with full harmonic content using templates with the (2,2) harmonic only. 
} 


\subsection{Looking at a subset of parameters}

In the expression we have derived, $n_p$ is the number of free parameters, i.e., the number of parameters that are varied when performing MCMC, and our criterion informs us if the true point lies inside the $n_p-{\rm dimensional}$ $90\%$ confidence region. Often, we are interested in estimating if biases are expected in a specific subset of parameters, e.g., the intrinsic parameters. We can denote $\theta=(\theta^1,\theta^2)$, where $\theta^1$ are the parameters we are interested in and $\theta^2$ those in which we are not. 

\subsubsection{Maximising approach}

A common approach is to maximise the overlap at the subset $\theta^1_0$ of the true point over the parameters $\theta^2$. We denote by $\check{\theta}^2$ the point that maximises $\mathcal{O}_T(s_0,h(\theta^1_0,\theta^2))$. Through Eq.~\ref{eq:log_like_olap}, this can be interpreted as maximising the likelihood over $\theta^2$ while fixing $\theta^1$ to $\theta^1_0$. With this procedure, we define a posterior distribution on $\theta^1$ that corresponds to a slice of the full posterior onto the $\theta^2=\check{\theta}^2$ hypersurface:
\begin{equation}
    p(\theta^1,\theta_2={\check{\theta}^2}|d_1,...,d_{n_d})=\frac{ p(\theta^1,{\check{\theta}^2}|d_1,...,d_{n_d})}{\int  p(\theta^1,{\check{\theta}^2}|d_1,...,d_{n_d}) {\rm d}\theta^1}. 
\end{equation}
It can be rewritten as:
\begin{equation}
     p(\theta^1,\theta_2={\check{\theta}^2}|d_1,...,d_{n_d})=\mathcal{N}\mathcal{L}(\theta^1,\check{\theta}^2),
\end{equation}
where $\mathcal{N}$ is a normalisation constant. If $\mathcal{L}(\theta^1,\theta^2)$ follows the Gaussian approximation, $\mathcal{L}(\theta^1,\check{\theta}^2)$ also does, and we can follow the same steps as before. We obtain the criterion:
\begin{equation}
     1-\mathcal{O}_T(s_0,h(\theta^1_0,\check{\theta}^2)) \leq \frac{n^1_p \left ( 1-\frac{2}{9 n^1_p}+1.3\sqrt{\frac{2}{9n^1_p}} \right )^3 }{2{\rm SNR}_T^2}+1-{\rm FF}_T(s_0,h(\check{\theta}^2)),\label{eq:crit_max_olap} 
\end{equation}
where $n^1_p$ are the number of parameters in the subset $\theta_1$, and ${\rm FF}_T(s_0,h(\check{\theta}^2))$ is the total fitting factor under the constraint $\theta^2=\check{\theta}^2$. 
The quantity $\mathcal{O}_T(s_0,h(\theta^1_0,\check{\theta}^2))$ is often called the faithfulness of the template. However, requiring $\theta^1_0$ to be in the $90\%$ confidence region of the distribution defined by fixing the parameters $\theta_2$ to the value $\check{\theta}^2$ in the posterior is not a well-motivated requirement. In particular, it is not the same as performing parameter estimation for all parameters $\theta$ and looking at the marginal posterior on $\theta^1$. Within the Gaussian approximation, this is true only if the sets of parameters $\theta^1$ and $\theta^2$ are uncorrelated, but in this case any value of $\theta^2$ could be used to evaluate the likelihood.

\subsubsection{Marginalising approach}\label{sec:marg_app}
From the Bayesian perspective, it is better defined to look at the marginalised posterior on $\theta^1$:
\begin{equation}
    p(\theta^1|d_1,...,d_{n_d})=\frac{\int \mathcal{L}(\theta^1,\theta^2)p(\theta^1,\theta^2) {\rm d}\theta^2}{\int \mathcal{L}(\theta^1,\theta^2)p(\theta^1,\theta^2) {\rm d}\theta^2{\rm d}\theta^1}. 
\end{equation}
In the case of flat priors, the prior term is constant, and the posterior on $\theta^1$ can be written as:
\begin{equation}
    p(\theta^1|d_1,...,d_{n_d})=\mathcal{N}\mathcal{L}_{\theta^2}(\theta^1),
\end{equation}
where we have introduced $\mathcal{L}_{\theta^2}$ as the average likelihood over the prior range on $\theta^2$. If the Gaussian approximation applies to the full likelihood, it still applies to the marginalised one. Thus, following the same reasoning as we did before, we obtain the indistinguishability criterion for $\theta^1$:
\begin{equation}
   \ln \mathcal{L}_{\theta^2}(\theta^1_0) \geq \ln \hat{\mathcal{L}_{\theta^2}}-\frac{\mathcal{Q}_{f}(n_p^1)}{2}.\label{eq:marg_crit}
\end{equation}
For the sake of comparison, we introduce the "averaged" overlap through:
\begin{equation}
    \ln \mathcal{L}_{\theta^2}(\theta^1)=-{\rm SNR_T}^2(1-\mathcal{O}_{T,\theta^2}(\theta^1)),
\end{equation}
and define the ``averaged'' fitting factor ${\rm FF}_{T,\theta^2}$ as the maximum of the ``averaged'' overlap over $\theta_1$.
The criterion then reads:
\begin{equation}
   1- \mathcal{O}_{T,\theta^2}(\theta^1_0 )\leq   \frac{n^1_p \left ( 1-\frac{2}{9 n^1_p}+1.3\sqrt{\frac{2}{9n^1_p}} \right )^3 }{2{\rm SNR}_T^2} + 1-{\rm FF}_{T,\theta^2}.
\end{equation}
Unfortunately, because of the averaging over $\theta^2$, this version of the indistinguishability criterion is not very convenient to work with. We now propose a different criterion to address the presence of biases within a subset of parameters.

\section{Model selection}

In~\cite{Toubiana:2023cwr}, we introduced the Akaike information criterion (AIC) \citep{1100705} of a model, defined as:
\begin{equation}
    {\rm AIC}=k\,n_p-2\ln\hat{\mathcal{L}}, \label{eq:aic}
\end{equation}
where $k$ is a constant, that was originally taken to be $k=2$, but other choices are occasionally made in the statistics literature. The
Bayes' factor between two models can be approximated by:
\begin{equation}
    \ln \mathcal{B}=-\frac{1}{2}({\rm AIC}_1-{\rm AIC}_2). \label{eq:log_bayes}
\end{equation}
This can be justified by considering the case of an $n_p-{\rm dimensional}$ Gaussian likelihood with covariance eigenvalues $\sigma_i$, and taking a flat prior that extends a width $2\Delta_i$ in each eigenvector direction, $\vec\theta_i$. The log-evidence for this model is: $\ln\left(\prod_i^{n_p}\frac{\sqrt{2\pi}\sigma_i}{2\Delta_i}\right)+\ln\hat{\mathcal{L}}$. Eq.~\ref{eq:log_bayes} is recovered by choosing $\Delta_i=\frac{e\sqrt{2\pi}}{2}\sigma_i\simeq 3.4 \sigma_i$. In this way, the chosen prior domain contains $0.9993^{n_p}$ of the likelihood probability weight, which is usually enough for the values of $n_p$ we are typically interested in. For spinning binary GW sources, $n_p$ can be as large as $15$, in which case this probability weight is $\sim99\%$. Increasing $\Delta_i$ to $4.07\sigma_i$ (which would mean that the prior now contains $99.93\%$ of the likelihood probability weight when $n_p=15$) would correspond to setting $k=2.36$ in the AIC. These choices of $\Delta_i$ are somewhat ad-hoc, but the procedure can be interpreted as trying to maximize the evidence for a model with flat priors, without excluding any region of the parameter space consistent with the observed data. It is interesting that the AIC can be related to the evidence in this way, since the AIC is derived using information theory, by minimizing the information loss relative to the true data generating process by considering the Kullback-Liebler divergence between different models. Moreover, note that the AIC is a measure of how well a model fits the data: the smaller it is, the better the fit. In this sense, interpreting the difference in AICs as a log-Bayes' factor through Eq.~\ref{eq:log_bayes} is a way of a introducing a scale to quantify how much a model is preferred to another.

Now, we discuss how this approach can be applied to assess if biases are expected. From the Bayesian perspective, we expect not to have biases in a subset of parameters $\theta^1$, if the model where those parameters are fixed to their true value $\theta^1_0$ is 
not disfavoured over the model where we vary them. This is the same reasoning we used in~\cite{Toubiana:2023cwr} to estimate the SNR at which we would expect apparent deviations from GR to arise due to systematic effects. The log-Bayes' factor between the two models reads:
\begin{equation}
    \ln \mathcal{B}=\ln \hat{\mathcal{L}}(\theta^1=\theta^1_0)-\ln \hat{\mathcal{L}}+ n^1_p.\label{eq:bias_bayes} 
\end{equation}
In that equation, $\ln \hat{\mathcal{L}}(\theta^1=\theta^1_0)$ is the maximum log-likelihood when fixing $\theta^1$ to $\theta^1_0$ and letting only $\theta^2$ vary, and $\ln \hat{\mathcal{L}}$ is the maximum log-likelihood when varying all the parameters. Using Eq.~\ref{eq:log_like_olap}, the log-Bayes' factor can be written:
\begin{equation}
    \ln \mathcal{B}=\left [ \mathcal{O}_T(s_0,h(\theta^1_0,\check{\theta}^2))-{\rm FF}_T(s_0,h) \right ]{\rm SNR}_T^2+ n^1_p,\label{eq:bias_bayes2} 
\end{equation}
where, as above, $\check{\theta}^2$ is such that the total overlap at $\theta^1_0$ is maximised.

The value of $ \ln \mathcal{B}$ from which we expect to observe biases is not unequivocally defined. Adopting the Kass-Raftery scale \cite{doi:10.1080/01621459.1995.10476572}, we expect to have strong evidence for a bias if $\ln \mathcal{B}$ is in the range 3-5, and very strong evidence for $\ln \mathcal{B}>5$. In~\cite{Toubiana:2023cwr}, we found these scales to be in good agreement with our parameter estimation runs.


\section{Practical considerations}\label{sec:prat_cons}
%
%

 We are often interested in determining the SNR threshold at which we expect biases to begin to appear. From Eq.~\ref{eq:logl_exp}, when varying the SNR of the source just by varying the distance of the signal $s_0$, the log-likelihood can be readily rescaled: since only the $(s_{0,i}|s_{0,i})$ term depends on the distance, each term of the sum scales as ${\rm SNR}_i^2$, and each of these scales in the same way with distance. Moreover, the parameters that maximise the log-likelihood do not change, apart from the distance, which scales in the same way as the distance of $s_0$. Thanks to this observation, the Bayes' factor-based criterion, as well as the full and maximised versions of the revisited indistinguishability criterion (given by Eqs.~\ref{eq:crit_general}, \ref{eq:crit} and \ref{eq:crit_max_olap}, not the one where we perform marginalisation over a subset of parameters presented in Sec.~\ref{sec:marg_app}) can be easily evaluated at different SNRs, using minimisation routines or Markov-Chain-Monte-Carlo (MCMC) (as in~\cite{Toubiana:2023cwr}) to estimate $\hat{\theta}^2$ and $\ln \hat{\mathcal{L}}$ for a given SNR and applying the proper rescaling. 
Alternatively, using the linear signal approximation, we can compute the Fisher matrix (see Eq.~\ref{eq:Fisher}), assuming that $\hat{\theta}\sim \theta_0$ (the difference in the Fisher matrix at those points should yield higher order corrections to the linear signal approximation), use it to draw samples of $\theta$ and compute the log-likelihood at those points. A drawback of this method and of the MCMC-based one, is that the maximum log-likelihood lies in the higher-end tail of the log-likelihood distribution, which might not be well sampled 
(recall that errors in the estimation of the log-likelihood will scale with ${\rm SNR}^2$ in the criteria). 
We can once again exploit the fact that the log-likelihood follows a $\chi^2(n_p)$ distribution and estimate its maximum value from the mean, $<\ln \mathcal{L}>$, whose estimate through sampling is more stable, using the relation:
 \begin{equation}
     \ln \hat{\mathcal{L}}=<\ln \mathcal{L}>+\frac{n_p}{2}.
 \end{equation}

The fact that the marginalised version of the revisited indistinguishability criterion (Eq.~\ref{eq:marg_crit}) does not follow the ${\rm SNR}$ scaling makes it even more inconvenient to work with. Thus, we believe the Bayes' factor-based criterion we propose offers a powerful alternative for the study of systematic effects.


\newpage
\section{Acknowledgements}

We are thankful to Lorenzo Pompili and Alessandra Buonanno for fruitful discussions that led to the elaboration of these notes. 

\bibliography{ref.bib}{}
\bibliographystyle{aasjournal}

\end{document}